\documentclass[sigconf, nonacm]{acmart}

\usepackage{amsmath}
\newcommand{\halfarrow}{\overset{\rightharpoonup}}

\usepackage{subcaption}
\usepackage[noabbrev,capitalize]{cleveref}
\Crefformat{figure}{#2Figure~#1#3}
\Crefformat{table}{#2Table~#1#3}
\usepackage{fancyhdr}
\AtBeginDocument{}

\setcopyright{acmlicensed}
\copyrightyear{2025}
\acmYear{2025}
\acmDOI{XXXXXXX.XXXXXXX}

\acmConference[ASP-DAC 2025]{Make sure to enter the correct
  conference title from your rights confirmation emai}{Jan. 20--23,
  2025}{Tokyo, Japan}
  
\acmISBN{978-1-4503-XXXX-X/18/06}

\fancypagestyle{mystyle}{
  \fancyhf{} 
  \fancyhead[R]{%
    \begin{tabular}{r}
        \textbf{\fontsize{12}{12}\selectfont Personal Data (D)} \\
        \textbf{\fontsize{12}{12}\selectfont Macronix Confidential}
    \end{tabular}
  }
}
\begin{document}

\title{Efficient and Reliable Vector Similarity Search Using Asymmetric Encoding with NAND-Flash for Many-Class Few-Shot Learning}

\author{Hao-Wei Chiang}
\affiliation{
  \institution{Graduate Institute of Electronics Engineering, National Taiwan University}
  \city{Taipei}
  \country{Taiwan}
}
\email{jackson@access.ee.ntu.edu.tw}

\author{Chi-Tse Huang}
\affiliation{
  \institution{Graduate Institute of Electronics Engineering, National Taiwan University}
  \city{Taipei}
  \country{Taiwan}
}
\email{rickhuang@access.ee.ntu.edu.tw}

\author{Hsiang-Yun Cheng}
\affiliation{
  \institution{Research Center for Information Technology Innovation, Academia Sinica}
  \city{Taipei}
  \country{Taiwan}
}
\email{hycheng@citi.sinica.edu.tw}

\author{Po-Hao Tseng}
\affiliation{
  \institution{Macronix International Co. Ltd.}
  \streetaddress{16 Li-Hsin Road, Hsinchu Science Park}
  \city{Hsinchu}
  \country{Taiwan}
}
\email{pohaotseng@mxic.com.tw}

\author{Ming-Hsiu Lee}
\affiliation{
  \institution{Macronix International Co. Ltd.}
  \streetaddress{16 Li-Hsin Road, Hsinchu Science Park}
  \city{Hsinchu}
  \country{Taiwan}
}
\email{Ericlee@mxic.com.tw}

\author{An-Yeu (Andy) Wu}
\affiliation{
  \institution{Graduate Institute of Electronics Engineering, National Taiwan University}
  \city{Taipei}
  \country{Taiwan}
}
\email{andywu@ntu.edu.tw}







\renewcommand{\shortauthors}{}

\renewcommand{\shorttitle}{}

\begin{abstract}
While memory-augmented neural networks (MANNs) offer an effective solution for few-shot learning (FSL) by integrating deep neural networks with external memory, the capacity requirements and energy overhead of data movement become enormous due to the large number of support vectors in many-class FSL scenarios. Various in-memory search solutions have emerged to improve the energy efficiency of MANNs. NAND-based multi-bit content addressable memory (MCAM) is a promising option due to its high density and large capacity. Despite its potential, MCAM faces limitations such as a restricted number of word lines, limited quantization levels, and non-ideal effects like varying string currents and bottleneck effects, which lead to significant accuracy drops. To address these issues, we propose several innovative methods. First, the Multi-bit Thermometer Code (MTMC) leverages the extensive capacity of MCAM to enhance vector precision using cumulative encoding rules, thereby mitigating the bottleneck effect. Second, the Asymmetric vector similarity search (AVSS) reduces the precision of the query vector while maintaining that of the support vectors, thereby minimizing the search iterations and improving efficiency in many-class scenarios. Finally, the Hardware-Aware Training (HAT) method optimizes controller training by modeling the hardware characteristics of MCAM, thus enhancing the reliability of the system. Our integrated framework reduces search iterations by up to $32\times$, and increases overall accuracy by $1.58\%$ to $6.94\%$.
\end{abstract}

\maketitle
\vspace{-1em}
\section{Introduction}

In recent years, memory-augmented neural networks (MANNs) \cite{ref::google_mann, ref::matching_networks, ref::fefet_tcam_mann} have gained significant traction, particularly in the domain of few-shot learning (FSL). MANN is composed of a feature extraction model (controller) to transform images into vector representations and an external memory module for storing vectors derived from a limited set of labeled samples known as the support set. During inference, the query image is also converted to query vector through controller and the prediction is made by retrieving these support vectors from external memory and comparing them with the query vector through vector similarity search (VSS). Recent studies \cite{ref::many_cls_maninet, ref::large_cls} have extended the use of MANNs to address many-class scenarios of few-shot learning, as they are more practical and prevalent in various real-world applications such as robot navigation\cite{ref::robot_fewshot}, medical imaging\cite{ref::medical_fewshot}, and video surveillance\cite{ref::surveillance_fewshot}. However, one of the primary challenges in this context is the substantial memory capacity required to store numerous vectors for all classes. Additionally, performing VSS necessitates frequent off-chip memory access to calculate the similarity between query and support vectors. This frequent memory access induces significant energy overhead in conventional von Neumann-based computing systems \cite{ref::ice_mann, ref::x_mann}. In many-class scenarios, this problem is exacerbated by the large number of support vectors for each class, further increasing the energy overhead of data movement.

\begin{figure}[t]
  \centering
  \includegraphics[width=0.9\linewidth]{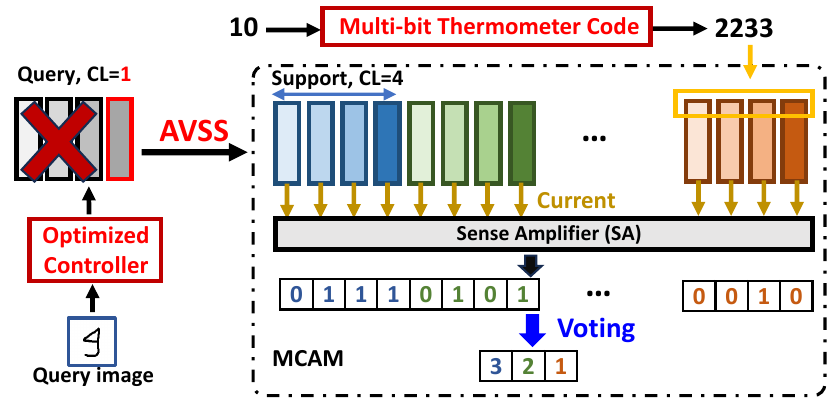}
  \vspace{-1em}
  \caption{The proposed processing flow using MCAM}
    \vspace{-2em}
    \label{fig::vss_voting_flow}
    \Description{}
\end{figure}

To address these issues, researchers have introduced in-memory search (IMS) \cite{ref::sapiens, ref::one_shot_learning_mann, ref::searchd}. This approach aims to reduce excessive vector movement and accelerate VSS by enabling parallel searches within content-addressable memories (CAMs). Among various memory devices, NAND flash stands out as particularly promising due to its ultra-high density and capacity, making it ideal for efficiently storing and managing large volumes of support vectors in the many-class FSL. Tseng \textit{et al.} proposed a 3D NAND-based multi-bit CAM (MCAM) \cite{ref::mxic_mcam}, capable of storing up to 128K vectors with 24 dimension. In MCAM, all support vectors are stored in memory, and the query vector is applied via the word line, with string (bit line) currents representing the similarity of each query-support vector pair. Higher string current can be measured for the query-support vector pair with higher similarity. Instead of using energy-consuming circuit to identify the accurate current value, a sensing amplifier (SA) with a voting scheme are applied to obtain the most similar support vector stored in the memory.

Despite its innovative design, the MCAM faces significant challenges.  First, each multi-level cell (MLC) in the MCAM can only represent four distinct states. If each dimension of the support vectors is directly mapped on a unit cell, the quantization level of support vectors is limited to 4. Previous study \cite{ref::camasim} shows that such limitation may significantly hurt the accuracy of MANN in classification tasks. 
Second, due to the serially connected architecture of NAND-based MCAM, the string current is influenced not only by the similarity between the stored vector and the search data but also by the cell with the lowest gate overdrive (largest cell mismatch level) within the NAND string. Consequently, the string current generated by a highly similar query-support vector pair may be significantly lower than that of a less similar pair if a large mismatch exists in any dimension of the similar pair. This bottleneck effect can lead to inconsistent VSS outcomes and negatively impact the accuracy of MANNs.
In addition, the limited number of unit cells per string can necessitate additional iterations in the VSS process when dealing with high-precision input vectors. This affects both the system’s parallelism and overall hardware utilization. Furthermore, some non-ideal effects such as device variation may lead to the large fluctuations of the measured string current, which greatly undermines the accuracy of MANNs.

As shown in \Cref{fig::vss_voting_flow}, this paper proposes three methods to overcome the aforementioned challenges of MCAM. First, rather than adopting the common bit-slicing approach such as base-4 encoding (B4E), we propose multi-bit thermometer code (MTMC) to increase the quantization level of vectors. MTMC adopts a cumulative encoding rule to enhance the precision and mitigate the bottleneck effect of NAND-based MCAM. Second, asymmetric vector similarity search (AVSS) is proposed to minimize the searching iterations attributed to the limited unit cells in each string of MCAM. Furthermore, we propose a hardware-aware training (HAT) mechanism to address the non-ideal effects of NAND-based MCAM. HAT models the behavior of NAND-based MCAM’s hardware characteristics during controller training, thereby improving the robustness of the controller. Combining these approaches, we reduces search iterations by up to $32\times$, and increases overall accuracy by $1.58\%$ to $6.94\%$. The key contributions of this work are as follows:

\vspace{-1em}
\begin{enumerate}
    \item \textbf{Multi-bit thermometer code for mitigating the bottleneck effect of MCAM and achieving higher precision: } Compared to the encoding method in prior works \cite{ref::sapiens, ref::b4_weighted, ref::b4_encoding}, MTMC improves the accuracy by $0.34\%$ to $4.91\%$ with the same energy consumption.
    \item \textbf{Asymmetric vector similarity search for reducing the searching iterations in many-class scenarios:} With AVSS, the search iterations of VSS can be reduced by $32\times$ and $25\times$ for the Omniglot \cite{ref::omniglot_dataset} and CUB \cite{ref::cub_dataset} dataset.
    \item \textbf{Hardware-aware training mechanism for controller optimization:} Modeling the hardware behavior in training through HAT further improves the accuracy by $1.25\%$ to $1.8\%$ compared to standard training method with MTMC.
\end{enumerate}

\vspace{-1em}
\section{Background and Motivation}
\subsection{Memory-Augmented Neural Network (MANN) for Few-Shot Learning (FSL)}
\label{sec::MANN_for_FSL}



In few-shot classification tasks, the training and testing datasets contain completely different classes. The N-way K-shot setting involves N different classes with K labeled examples for each class. A key challenge in FSL is how to utilize limited data to adapt to unseen classes effectively. MANN offers a viable solution by enhancing deep neural networks (DNNs) with external memory. By leveraging this memory to retain encountered data, MANN can efficiently employ vector similarity search (VSS) to calculate similarities and make predictions for new tasks using a few support data.


The process of VSS is crucial in MANNs. One of the most widely used metrics for evaluating the similarity between query and support vectors is cosine similarity. It measures the cosine of the angle between two vectors, providing a value between -1 and 1. A value of 1 (maximum similarity) indicates that the vectors have identical orientations, while -1 (minimum similarity) indicates that they are completely opposite in orientation.
This metric has proven effective due to its sensitivity to the orientation of the vectors, making it prominent in various applications. However, implementing cosine similarity in IMS systems is challenging due to the complexity of the operations involved. As alternatives, researchers have proposed several hardware-efficient distance metrics such as $L_1$\cite{ref::sapiens}, $L_2$\cite{ref::l2_mcam}, and $L_{\infty}$\cite{ref::bore} that leverage the functionality of content addressable memory (CAM) to perform VSS directly in memory, significantly reducing the overhead of data movement.

\vspace{-1em}
\subsection{Mult-bit Content Addressable Memory}
\label{sec::mcam_background}
\begin{figure}[t]
    \centering
    \begin{subfigure}[b]{0.6\linewidth}
        \centering
        \includegraphics[width=\linewidth]{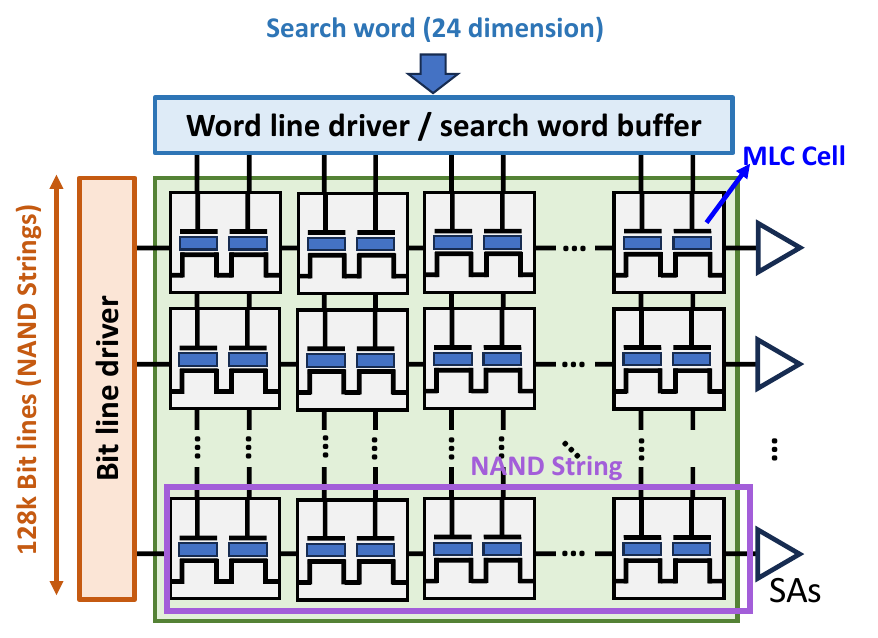}
        \vspace{-1em}
        \caption{}
        \label{fig::mcam_schematic}
    \end{subfigure}
     \begin{subfigure}[b]{0.38\linewidth}
        \begin{subfigure}[b]{\linewidth}
             \centering
            \includegraphics[width=\linewidth]{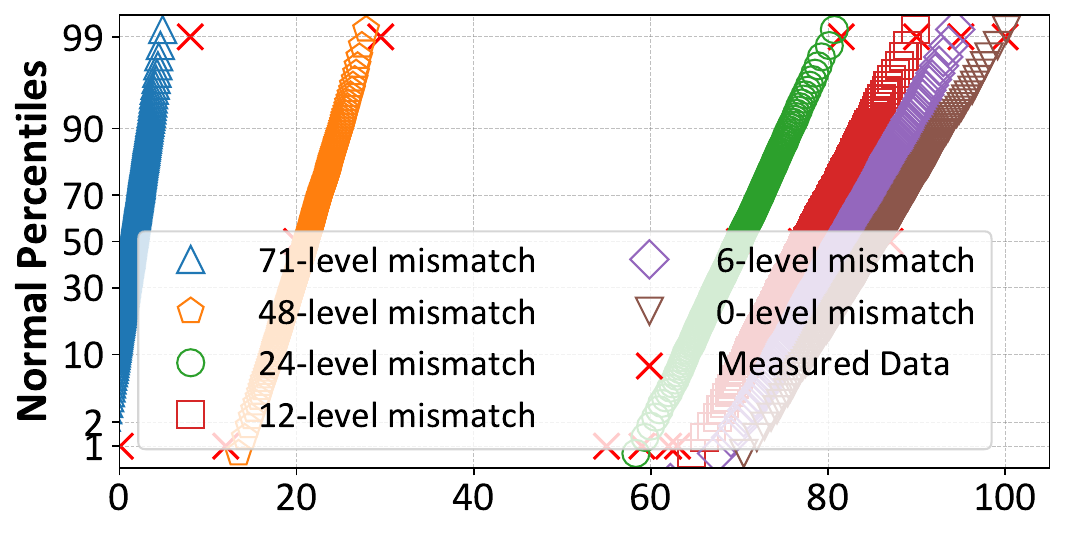}
            \vspace{-2em}
            \caption{}
            \label{fig::mcam_current}
        \end{subfigure}
        \begin{subfigure}[b]{\linewidth}
            \centering
            \includegraphics[width=\linewidth]{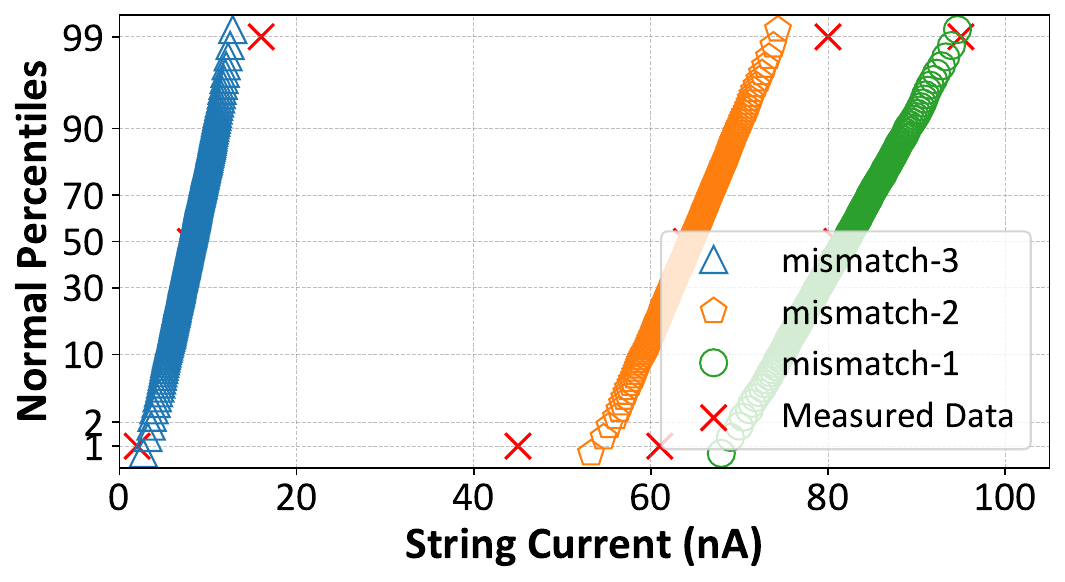}
            \vspace{-2em}
            \caption{}
            \label{fig::mcam_bottleneck}
        \end{subfigure}
    \end{subfigure}
    \vspace{-1em}
    \caption{(a) The Schematic of MCAM. (b) Simulated current distributions of MCAM with various string mismatch level. (c) Simulated current distributions of MCAM with 6-level string mismatch level, but with different maximum mismatch level in each string. The measured data points in (b) and (c) are derived from \cite{ref::mxic_mcam}.}
    \Description{}
    \vspace{-2em}
\end{figure}

To perform VSS with the $L_1$ metric, as described in \Cref{equ::l1}, directly in flash memory, Tseng \textit{et al.} proposed an in-memory approximate search (IMAS) system \cite{ref::mxic_mcam}.
\begin{equation}
    \label{equ::l1}
    L_1(\halfarrow{q}, \halfarrow{s}) = \Sigma_i |{q_i} - {s_i}|
\end{equation}
The IMAS system supports approximate $L_1$ computation by integrating NAND-based Multi-bit Content Addressable Memory (MCAM) with peripheral circuits, including sense amplifiers (SAs). The schematic of NAND-based MCAM is shown in \Cref{fig::mcam_schematic}. Thanks to the high density of flash devices, a single block of NAND-based MCAM contains 128K NAND strings, each with 24 dimensions (unit cells). Input search voltages can be compared with up to 128k NAND strings within one cycle. The current of each NAND string represents the similarity of search data and the stored data. Higher current indicating higher similarity. By setting the desired current threshold of the SAs, the NAND strings with current larger than the threshold can be efficiently identified.

The unit cell of MCAM consists of two serially connected MLC flash devices. The search current of a unit cell is determined by the gate overdrive levels between the search bias of the word lines and the threshold voltage of the flash device in the unit cell. According to the encoding scheme mentioned in \cite{ref::mxic_mcam}, each unit cell (dimension) can produce four possible matching results, ranging from 0-level mismatch (mismatch-0) to 3-level mismatch (mismatch-3). The matching current of each NAND string is determined by the string mismatch level, which is the sum of the mismatch levels of each dimension within the NAND string. The string mismatch level ranges from 0-level mismatch to 72-level mismatch in the 48-layer NAND strings. As shown in \Cref{fig::mcam_current}, larger mismatch levels in a NAND string result in smaller currents, effectively representing the similarity between the search and stored vectors. However, there is some variation in the matching current due to device variation. Additionally, the current flowing through the NAND string is influenced by the cell with the maximum mismatch level. This bottleneck effect means that even if most cells have low mismatch levels, a single cell with a maximum mismatch level can significantly reduce the overall current. We use mismatch-$n$ to denote the maximum mismatch levels. As shown in \Cref{fig::mcam_bottleneck}, even under the same total mismatch levels, the string with mismatch-3 (e.g., search data is 0 and stored data is 3) shows the smallest current, while the string with mismatch-1 leads to the largest current.


\begin{figure}[t]
    \centering
    \begin{subfigure}[b]{0.9\linewidth}
        \centering
        \includegraphics[width=\linewidth]{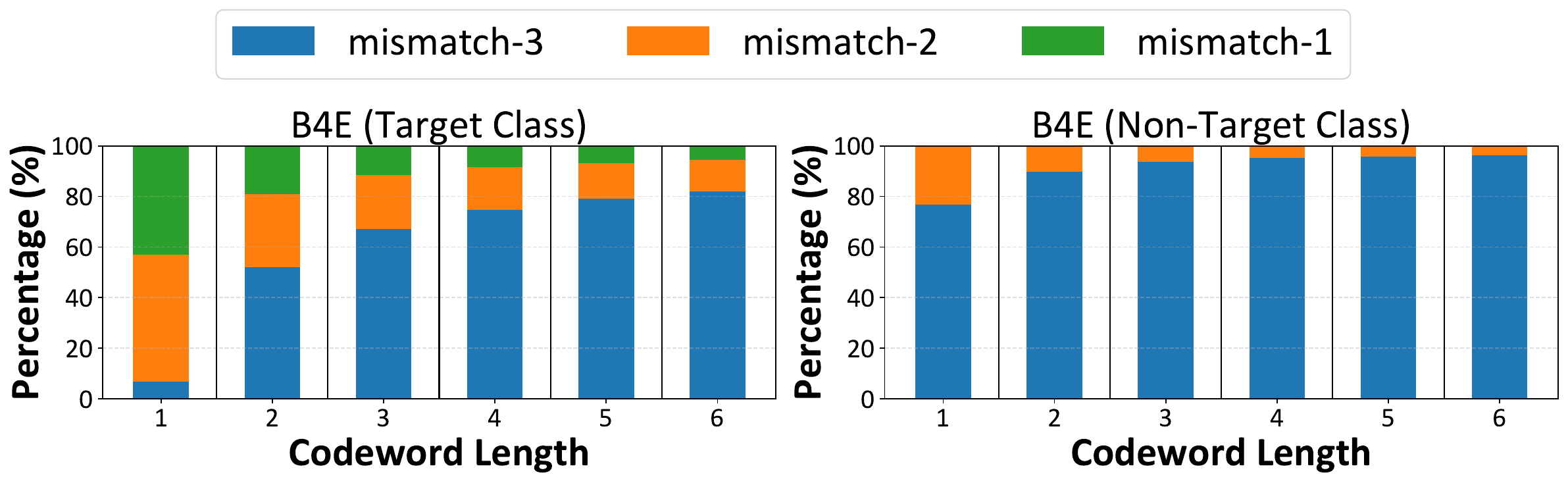}
        \vspace{-2.5em}
        \caption{}
        \label{fig::mismatch_percentage_base4}
    \end{subfigure}
    \hfill
    \begin{subfigure}[b]{0.9\linewidth}
        \centering
        \includegraphics[width=\linewidth]{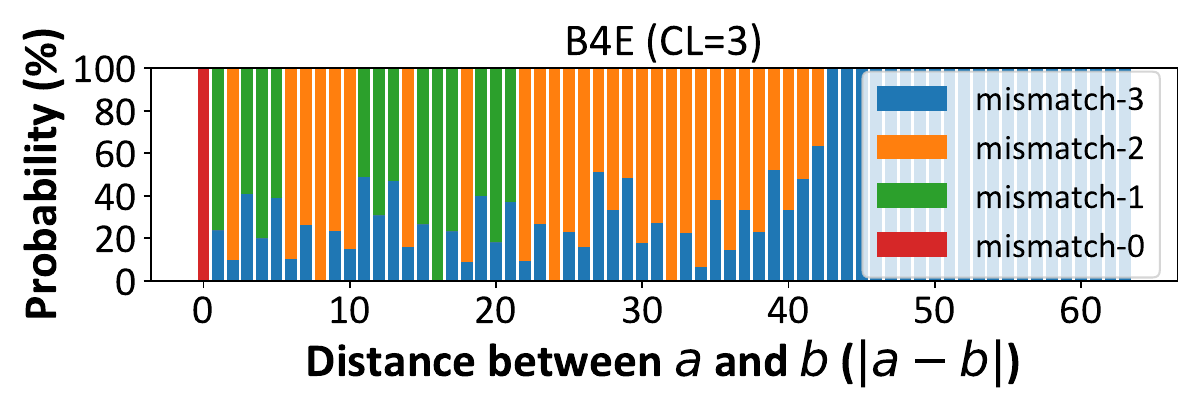}
        \vspace{-2.5em}
        \caption{}
        \label{fig::mismatch_distribution_base4}
        \vspace{-1em}
    \end{subfigure}
    \caption{(a) Distribution of each type of mismatch level with B4E (b) occurrence probability of each type of mismatch level under difference distance}
    \Description{}
    \vspace{-2em}
\end{figure}

\vspace{-1em}
\subsection{Motivation}


Applying the IMAS system with NAND-based MCAM to VSS in MANNs represents a promising advancement. The high density and large capacity of MCAM make it a prominent choice for many-class few-shot learning scenarios that require storing numerous vectors. However, several challenges must be addressed to effectively implement MCAM for VSS in MANNs, particularly for many-class few-shot learning scenarios. 


One primary challenge is that each unit cell in MCAM supports only 4 distinct levels of threshold voltage, which limits the quantization levels of vectors and is insufficient for many-class FSL. Based on our analysis, this limitation leads to $9.45\%$ accuracy degradation compared to a floating-point implementation on the Omniglot dataset. To increase precision of vectors, encoding techniques such as base-4 encoding (B4E) can be utilized. B4E encodes a value into multiple code words through bit-slicing (e.g., value 7 is encoded as 13 in B4E), which allows each code word to be mapped to a unit cell in MLC-based MCAM.

We analyzed the mismatch levels of all query and support vectors on the testing data of Omniglot dataset in \Cref{fig::mismatch_percentage_base4}. The target class includes query and support vectors that belong to the same class, while the non-target class includes vectors from different classes. Although B4E can enhance precision, it exacerbates bottleneck effects in NAND-based MCAM due to higher proportions of mismatch-3 as the code word length increases.
In addition, we evaluated the maximum mismatch level of all possible value pairs ($a, b$) under 64 quantization levels (a code word length $CL$ of 3) for B4E, where $a,b \in [0, 64]$. In \Cref{fig::mismatch_distribution_base4}, mismatch-3 may occur even when the distance between the $a$ and $b$ is small. This indicates that for similar query-support pairs encoded in B4E, the matching current and the computed similarity, expected to be high, can be significantly reduced due to the mismatch-3 occurrences in some dimensions. This results highlight the limitations of increasing precision through B4E.



\begin{figure}[t]
    \centering
    \begin{subfigure}[b]{0.9\linewidth}
        \centering
        \includegraphics[width=\linewidth]{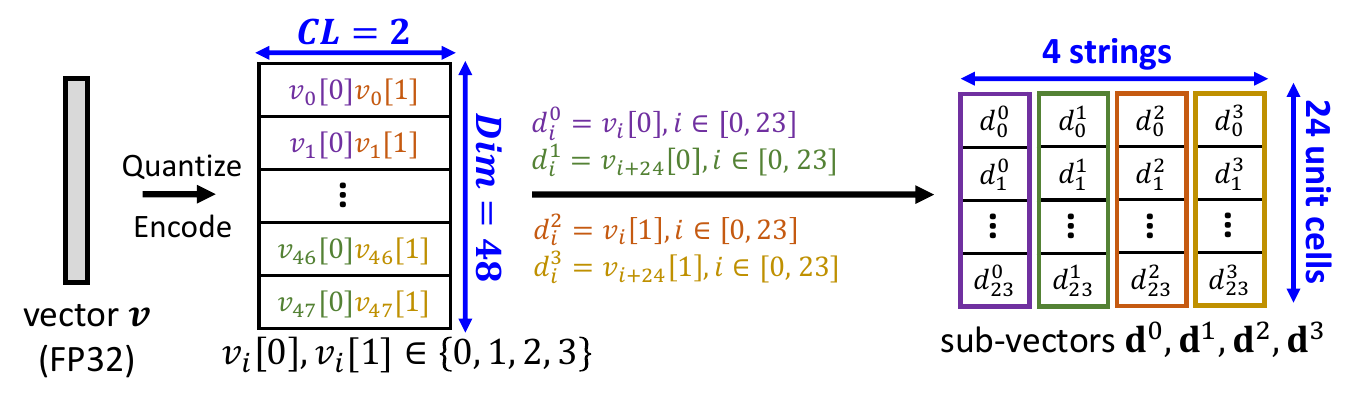}
        \vspace{-2.5em}
        \caption{}
        \label{fig::mcam_mapping}
    \end{subfigure}
    \hfill
    \begin{subfigure}[b]{0.85\linewidth}
        \centering
        \includegraphics[width=\linewidth]{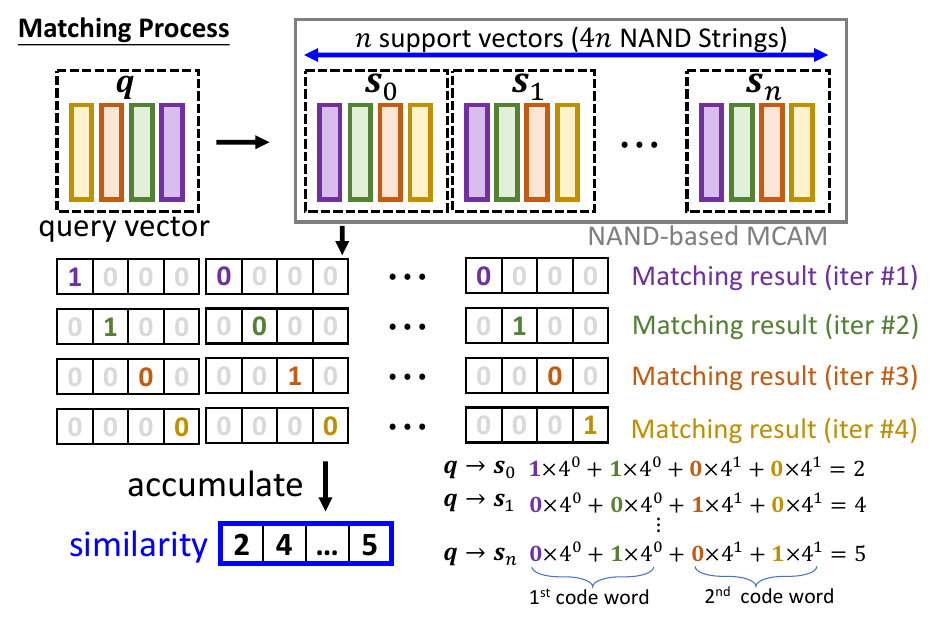}
        \vspace{-2.5em}
        \caption{}
        \label{fig::mcam_vss_detail}
    \end{subfigure}
    \vspace{-1em}
    \caption{Illustration of (a) mapping vectors with large dimensions or long code word length in MCAM (b) the matching process in MCAM.}
    \label{fig::mcam_detail}
    \Description{}
    \vspace{-2em}
\end{figure}
Second, the limited number of unit cells in each NAND string leads to more search iterations for input vectors with long code word lengths. In MCAM, each bit line has only 24 unit cells. For input vectors with the code word length larger than 24, the searching process requires multiple cycles. We employ a toy example to demonstrate the mapping method and VSS process of MCAM for this situation. As shown in \Cref{fig::mcam_mapping}, for a vector $\textbf{{v}}$ with 48 dimensions and a code word length of 2, it can be segmented into four sub-vectors $\textbf{d}^0$, $\textbf{d}^1$, $\textbf{d}^2$, and $\textbf{d}^3$. In the search process of MCAM, as shown in \Cref{fig::mcam_vss_detail}, it requires four iterations for each sub-vector of the query vector to be compared with the corresponding sub-vectors of all support vectors stored in NAND-based MCAM. The similarity of each query-support vector pair can be obtained from accumulating the matching result of each iteration. For B4E, the matching result ($mr_i$) of $i$-th code word is accumulated as shown in \Cref{equ::accumulate_matching_result}.
\begin{equation}
    \label{equ::accumulate_matching_result}
    similarity = \Sigma_i mr_i \times s_i, s_i = 4^{i-1}
\end{equation}
In general, for the vectors with dimension $d$ and code word length $CL$, the VSS process requires $k$ cycles, as each stored vector is split across $k$ adjacent NAND strings, where $k = \left \lceil{d \times CL/24}\right \rceil$.

Finally, non-ideal effects such as device variation and bottleneck effects significantly impact similarity measurements in VSS for MANNs. Device variations from both fabrication and write operations cause threshold voltage differences, leading to deviations in current levels and errors in similarity measurements. Bottleneck effects further degrade accuracy because the string current is restricted by the maximum mismatch level in a NAND string. Our analysis indicates these non-ideal effects result in over a $3.67\%$ accuracy loss on Omniglot dataset.

To address these challenges, it is crucial to enhance the encoding scheme to minimize the impacts of bottleneck effects and device variations, and to optimize the search flow to improve efficiency. Additionally, developing training algorithms specific to the characteristics of MCAM is vital. By integrating these techniques, we aim to improve the reliability and efficiency of VSS in MANNs, thus realizing the full potential of MCAM and significantly enhancing performance in many-class FSL scenarios.

\begin{table}
  \caption{Encoding rules of base-4 encoding (B4E) and multi-bit thermometer code (MTMC)}
  \vspace{-1em}
  \label{tab::mtmc_codebook}
  \begin{tabular}{ccc|ccc}
    \toprule
    Value & B4E & MTMC & Value & B4E & MTMC \\
    \midrule
    0 & 00 & 00000 & 8  & 20 & 11222 \\
    1 & 01 & 00001 & 9  & 21 & 12222 \\
    2 & 02 & 00011 & 10 & 22 & 22222 \\
    3 & 03 & 00111 & 11 & 23 & 22223 \\
    4 & 10 & 01111 & 12 & 30 & 22233 \\
    5 & 11 & 11111 & 13 & 31 & 22333 \\
    6 & 12 & 11112 & 14 & 32 & 23333 \\
    7 & 13 & 11122 & 15 & 33 & 33333 \\
    \bottomrule
  \end{tabular}
  \vspace{-1em}
\end{table}

\begin{figure}[t]
    \centering
    \begin{subfigure}[b]{0.9\linewidth}
        \centering
        \includegraphics[width=\linewidth]{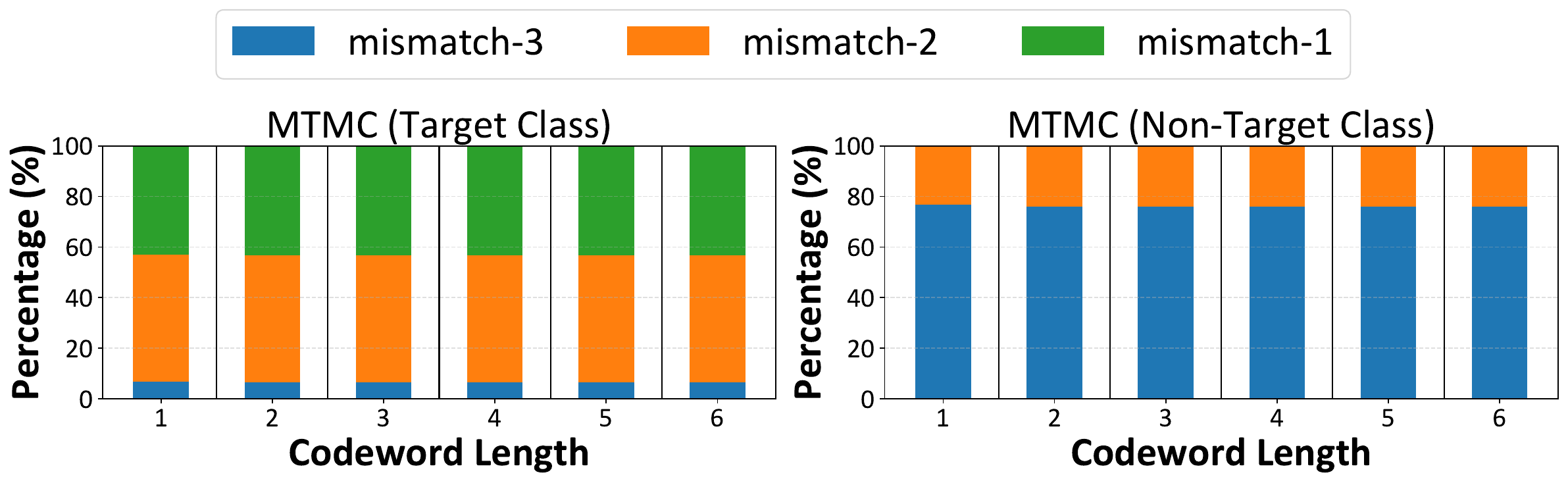}
        \vspace{-2em}
        \caption{}
        \label{fig::mismatch_percentage_mtmc}
    \end{subfigure}
    \hfill
    \begin{subfigure}[b]{0.8\linewidth}
        \centering
        \includegraphics[width=\linewidth]{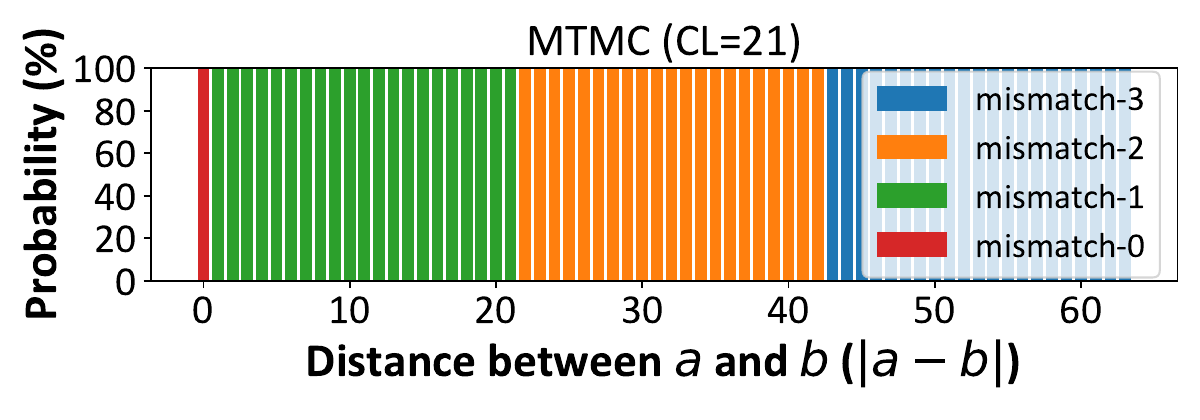}
        \vspace{-2em}
        \caption{}
        \label{fig::mismatch_distribution_mtmc}
    \end{subfigure}
    \vspace{-1em}
    \caption{(a) Distribution of each type of mismatch level with MTMC (b) occurrence probability of each type of mismatch level under difference distance}
    \Description{}
    \vspace{-1em}
\end{figure}

\section{Proposed Method}
In this section, we present the technical details of our work. \Cref{sec::MTMC} introduces the multi-bit thermometer code (MTMC) for enhancing the precision of VSS in MANNs by leveraging the capacity of MCAM. \Cref{sec::AVSS} presents asymmetric vector similarity search (AVSS) to increase the search parallelism of IMS. \Cref{sec::HAT} illustrates the hardware-aware training (HAT) technique, enabling the controller to make accurate predictions under non-ideal conditions combined with MTMC and AVSS.

\begin{figure}[t]
    \centering
    \begin{subfigure}[b]{0.45\linewidth}
        \centering
        \includegraphics[width=\linewidth]{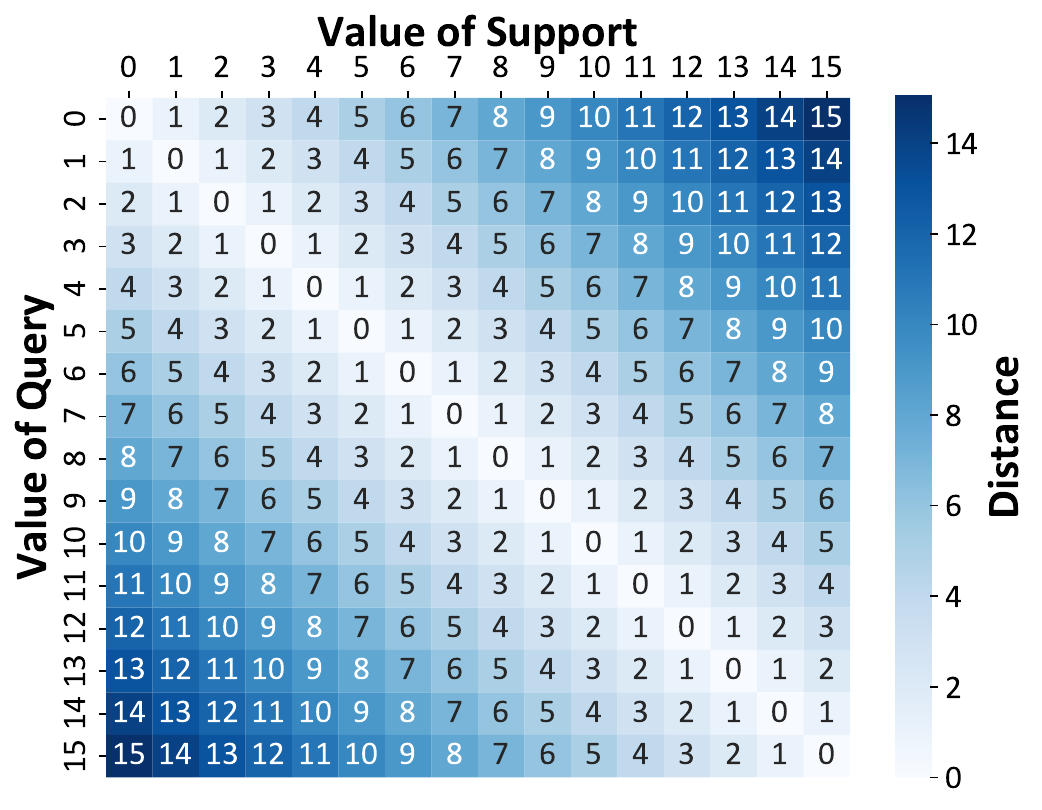}
        \vspace{-2em}
        \caption{}
    \end{subfigure}
     \begin{subfigure}[b]{0.45\linewidth}
        \centering
        \includegraphics[width=\linewidth]{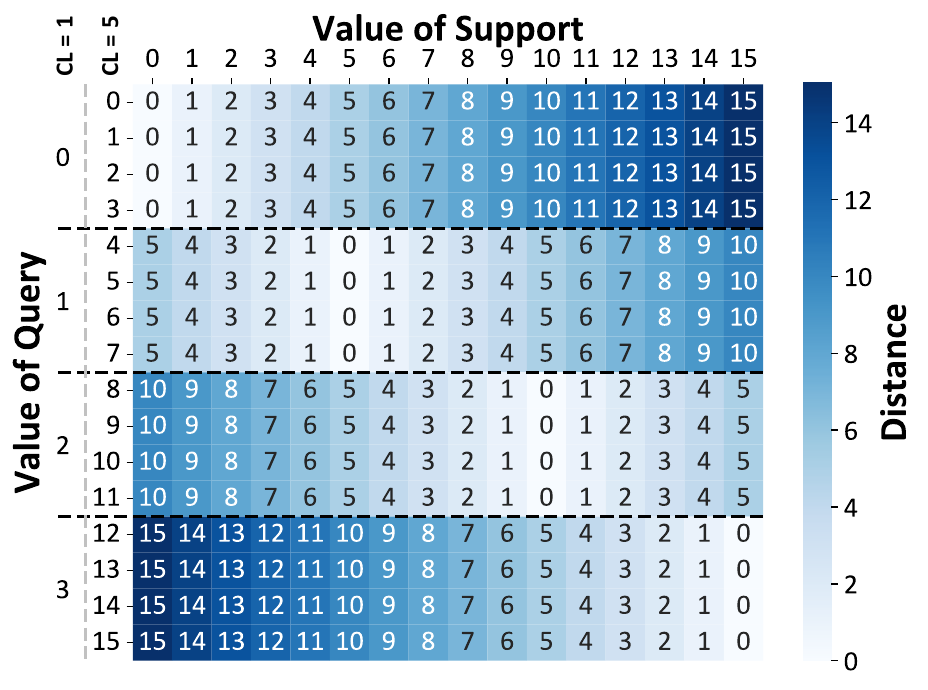}
        \vspace{-2em}
        \caption{}
    \end{subfigure}
    \vspace{-1em}
    \caption{Distance between query and support for (a) SVSS and (b) AVSS.}
    \label{fig::vss_heatmap}
    \Description{}
    \vspace{-2em}
\end{figure}

\begin{figure}[t]
  \centering
  \includegraphics[width=0.9\linewidth]{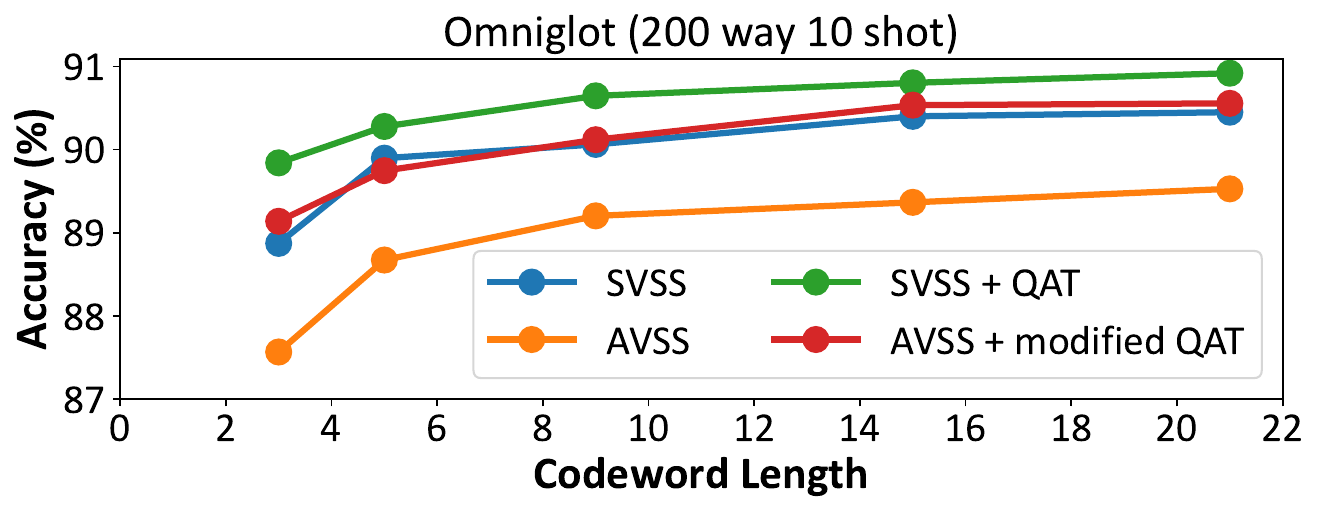}
  \vspace{-1em}
  \caption{Accuracy comparison between SVSS and AVSS before and after applying QAT.}
    \label{fig::vss_acc_comp}
    \Description{}
    \vspace{-2em}
\end{figure}



\vspace{-1em}
\subsection{Multi-bit Thermometer Code}
\label{sec::MTMC}

To enhance the precision of vectors and the reliability of VSS with MCAM, we propose multi-bit thermometer code (MTMC). This approach extends the traditional binary thermometer code into 4 levels per code word to leverage the 4 programmable states of MLC. The encoding rules for B4E and MTMC are listed in \cref{tab::mtmc_codebook}. Unlike B4E, where the actual value represented by each code word increases exponentially, the proposed MTMC represents values cumulatively. For a code word length $CL$, the value $m$ is represented by the first $CL-n$ code words set to $x$ and the remaining $n$ code words set to $x+1$, where $x=\left \lfloor{m/CL}\right \rfloor, n=mod(m, CL) $. This encoding ensures that the difference of each code word between two consecutive values is one, which is crucial for reducing errors caused by the bottleneck effect.

\begin{figure*}[t]
    \begin{subfigure}[b]{0.75\linewidth}
        \centering
        \includegraphics[width=\linewidth]{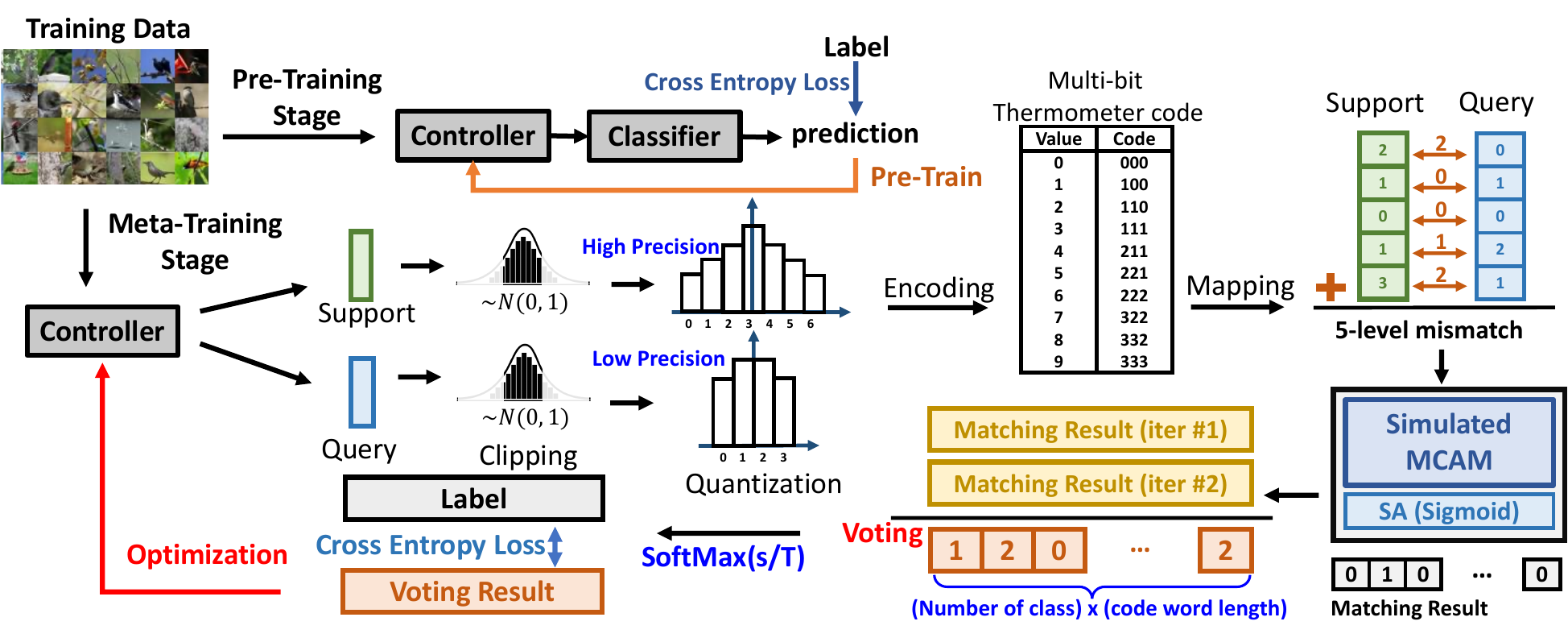}
        \vspace{-2em}
        \caption{}
    \end{subfigure}
    \hfill
    \begin{subfigure}{0.2\linewidth}
        \begin{subfigure}{\linewidth}
            \centering
            \includegraphics[width=\linewidth]{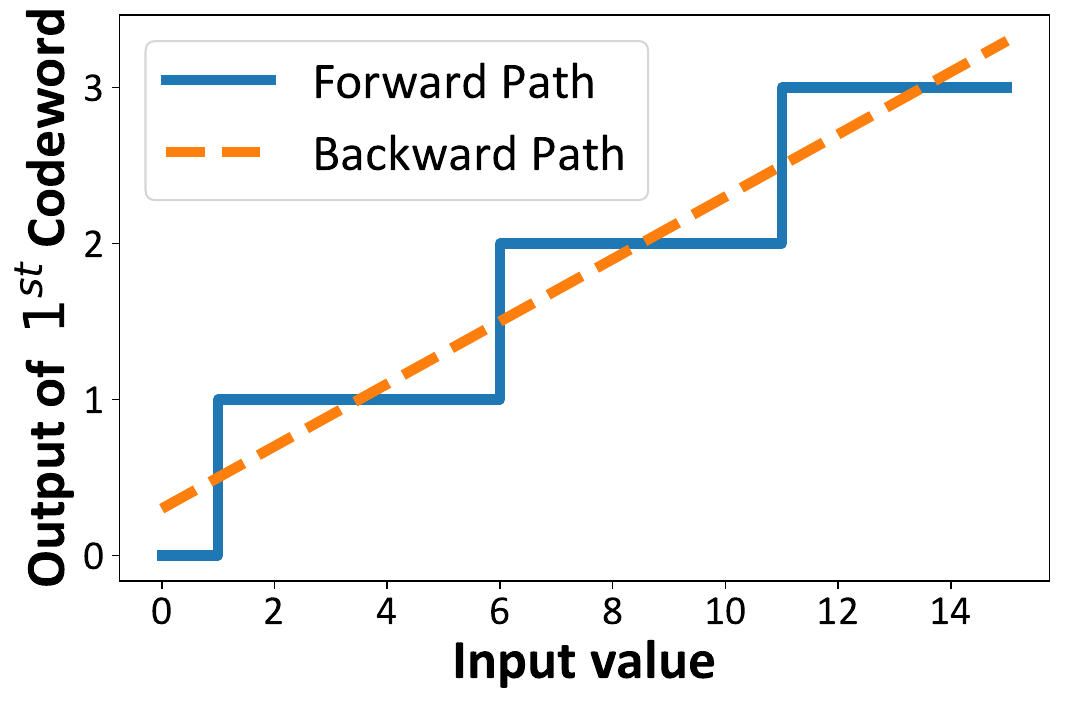}
            \vspace{-2em}
            \caption{}
            \label{fig::mtmc_gradient}
        \end{subfigure}
        \vfill
        \begin{subfigure}{\linewidth}
            \centering
            \includegraphics[width=\linewidth]{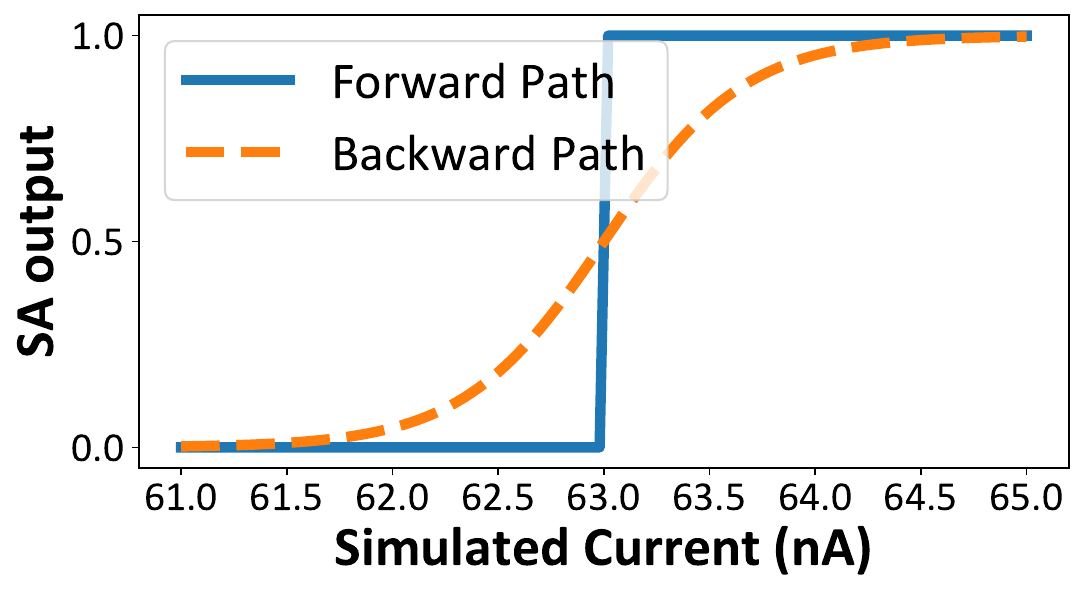}
            \vspace{-2em}
            \caption{}
            \label{fig::hat_sa}
        \end{subfigure}
    \end{subfigure}
  \centering
  \vspace{-1em}
  \caption{(a) Training flow of the proposed Hardware-aware training. (b) Forward and backward path of the encoding function for the $1^{st}$ code word of MTMC. (c) Forward and backward path of SA in simulated MTMC.}
  \label{fig::hat_flow}
  \Description{}
  \vspace{-1em}
\end{figure*}

To validate the effectiveness of MTMC in enhancing precision and mitigating the bottleneck effect, we conducted several analyses on mismatch levels. \Cref{fig::mismatch_percentage_mtmc} illustrates the percentage of each type of mismatch levels at various code word lengths while \Cref{fig::mismatch_distribution_mtmc} depicts the distribution of mismatch levels for MTMC. Comparing these results with those obtained using B4E (as shown in \Cref{fig::mismatch_percentage_base4} and \Cref{fig::mismatch_distribution_base4}), we observe that there is almost no changes in the percentage of mismatch-3 as the code word length increases when using MTMC, indicating that increasing the precision does not induce additional mismatch-3 and the errors introduced by the bottleneck effect are minimized. Furthermore, the mismatch distribution analysis in \Cref{fig::mismatch_distribution_mtmc} reveals that MTMC ensures that the maximum mismatch level of a value pair is small if the distance between the value pair is small. Specifically, for any pair of values with a code word length $CL$, only mismatch-0 or mismatch-1 may occur if the difference between these two values is less than $CL$. This property guarantees that the matching current of support vectors within the target class is significantly higher than that of support vectors in non-target classes, which facilitates a more reliable and accurate prediction in VSS.

\vspace{-2em}
\subsection{Asymmetric Vector Similarity Search}
\label{sec::AVSS}

When it comes to VSS process, the naive approach involves comparing the query and support vectors word-by-word, referring to as symmetric vector similarity search (SVSS) \cite{ref::sapiens}. While applying MTMC to vectors effectively increases precision, it also necessitates more comparisons for each query-support vector pair. As mentioned in \Cref{sec::mcam_background}, given that the length of NAND string in MCAM is limited to $24$, it requires $\left \lceil (CL\times d)/{24}\right \rceil $ adjacent NAND strings to store a support vector with a setting of $d$ dimensions and a code word length of $CL$. Since the NAND strings in MCAM share the same word line, $\left \lceil (CL\times d)/{24}\right \rceil $ iterations are also needed to compare each code word of the query and support in the VSS process, which significantly reduces the parallel advantage of performing VSS using NAND-based MCAM.

To address the latency overhead caused by longer code word length, we propose asymmetric vector similarity search (AVSS). AVSS sets the code word length of the query vector to 1, limiting the quantization level of the query vector to $4$. When performing AVSS, the single code word of the query vector is compared with all code words of the support vectors in the corresponding dimension. This approach reduces the required search iterations from $\left \lceil (CL\times d)/{24}\right \rceil $ to $\left \lceil{d/24}\right \rceil $ for vectors with dimension $d$. 


The limited quantization levels of the query vector in AVSS produce some unexpected incorrect distance measurements, as shown in \Cref{fig::vss_heatmap}. This limitation results in an approximate 1.5\% accuracy drop in MANNs, as demonstrated in \Cref{fig::vss_acc_comp}. To address this issue, we modified the quantized-aware training (QAT) technique from \cite{ref::qat} and trained the controller with different quantization schemes for the query and support vectors.
This training allows the controller to learn the effects of asymmetric quantization and the errors introduced by AVSS, and adjust its parameters accordingly. As a result, the accuracy gap between SVSS and AVSS narrows to within 1\%, as illustrated in \Cref{fig::vss_acc_comp}. These refinements in the VSS process highlight the potential of AVSS to reduce latency overhead while maintaining high accuracy, thereby optimizing the performance of MANNs in practical applications with NAND-based MCAM.

\vspace{-2em}
\subsection{Hardware-aware Training}
\label{sec::HAT}

The training flow of the proposed Hardware-Aware Training (HAT) mechanism, illustrated in \Cref{fig::hat_flow}, involves a two-stage process: pre-training and meta-training. In the pre-training stage, the controller is trained from scratch with a classifier to minimize the standard cross-entropy loss using all training samples. This widely adopted technique in few-shot learning methods \cite{ref::meta_baseline, ref::mixup_fewshot, ref::baseline_fewshot, ref::closer_fewshot} enables the model to learn robust and transferable feature representations from ample training data. During the meta-training stage, we incorporate hardware behaviors with episodic training to mimic the testing scenario. This allows the controller to learn how to extract feature vectors considering hardware behaviors and constraints.

At the beginning of the meta-training stage, both query and support images are transformed into vector representations through the pre-trained controller. Following the modified QAT method mentioned in \Cref{sec::AVSS}, query and support vectors are quantized into fixed-point values using different quantization schemes with 4 level quantization for query and $l$ level for support, where $l$ is a pre-defined parameter depending on the desired code word length of the support vectors. This enables the controller to adapt to the asymmetric setting of the quantization level for query and support, which is the condition it will encounter during testing. Besides, neural networks often produce outputs with a wide range of values, and the outliers can disproportionately affect the quantization process, leading to large information loss between floating-point and fixed-point values. Thus, the outputs of the controller are clipped within a range determined by the standard deviation of the outputs before quantization.

The quantized vectors are then encoded with the proposed MTMC. \Cref{fig::mtmc_gradient} illustrates the discrete encoding function mapping fixed-point value to the first bit of the MTMC code word. The encoding function is piece-wise constant, resulting in the zero gradient regardless of the input value and making the standard back-propagation process inapplicable.
However, we observed that the trend of the encoding function follows a line with a slope of $1/CL$. 
Inspired by the straight-through estimator approach \cite{ref::sta}, we estimate the gradient of the encoding function as a linear function during back propagation.

Furthermore, we built a simulated MCAM for analytical modeling the MCAM behavior when performing AVSS with the encoded query and support vectors. To minimize the accuracy degradation caused by the non-idealities of the MCAM, noise derived from gaussian distribution \cite{ref::camasim} is included in the simulated MCAM. For the behavior of SA in MCAM, directly using step function is not applicable since it is not continuous and the derivative of step function is a zero function. Thus, we use the gradient of sigmoid function to replace the gradient of step function during backward process, as depicted in \Cref{fig::hat_sa}. Finally, the voting result of the simulated CAM is used to calculate the cross-entropy loss to optimize the the controller. With the two-stage training flow, we can obtain a controller not only with superior generalization ability but also robust to non-ideality of MCAM.
\vspace{-1em}
\section{Evaluation and Results}
\subsection{Experimental Setup}
We conducted experiments on two few-shot learning (FSL) tasks to validate the efficacy of our design. For the Omniglot \cite{ref::omniglot_dataset} dataset, we utilized the Conv4 \cite{ref::matching_networks} architecture with 48 dimensions. This dataset comprises 1623 classes split into 964 classes for training and 659 for testing.  For the CUB-200-2011 \cite{ref::cub_dataset} (referred as CUB) dataset, we employed a more complex architecture, ResNet12 \cite{ref::resnet12_ref}, configured with 480 dimensions. Following the setting in \cite{ref::negative_margin}, the 200 classes are divided into 100 for training, 50 for validation, and 50 for testing. To adhere to the "many-class" scenario in our experiments, we adopted a 200-way 10-shot setting for Omniglot, which includes more classes than the setting in \cite{ref::robust_hd_mann}. Up to 128k NAND strings is needed under this setting with the code word length set to 32. As for CUB, a 50-way 5-shot setting, which occupied up to 125k NAND strings with the code word length of 25, was applied. We utilized the measurement results reported in previous research \cite{ref::mxic_mcam} to estimate the search energy required for our tasks.

\vspace{-1em}
\subsection{Pareto Front of Energy-Accuracy Trade-off}

\begin{figure}[t]
    \centering
    \begin{subfigure}[b]{0.48\linewidth}
        \centering
        \includegraphics[width=\linewidth]{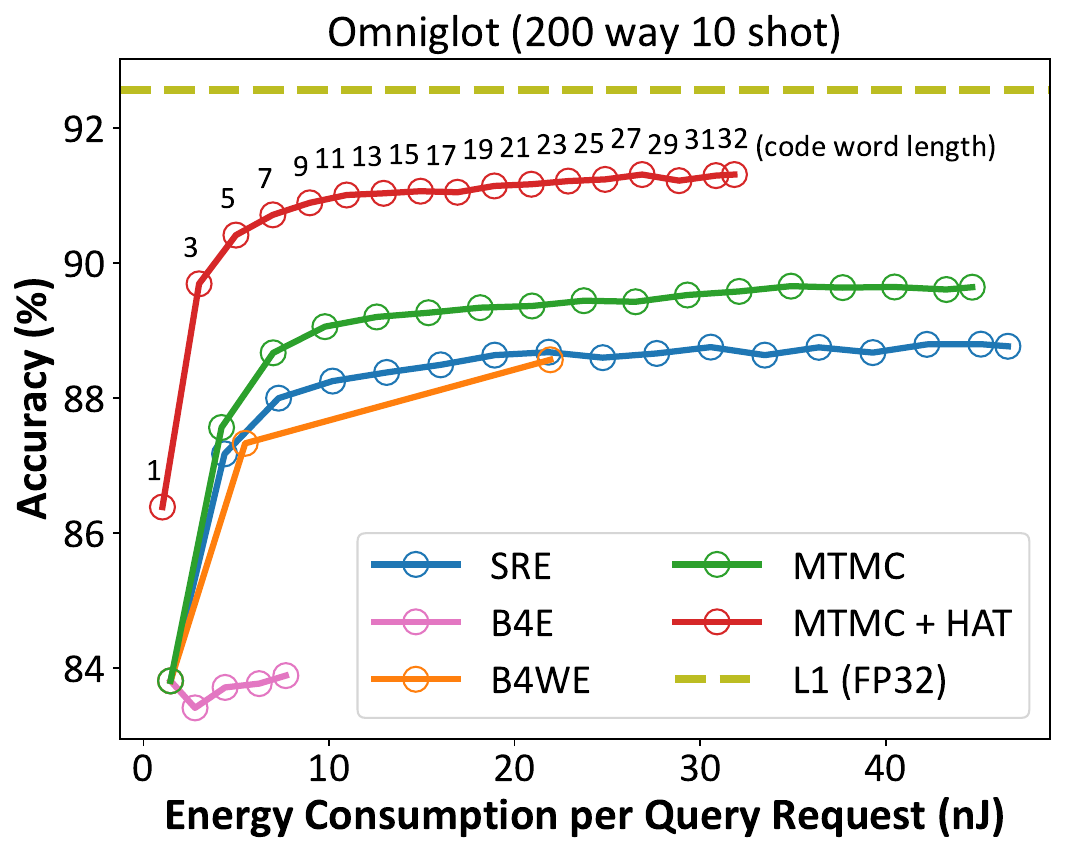}
        \vspace{-2em}
        \caption{}
    \end{subfigure}
     \begin{subfigure}[b]{0.48\linewidth}
        \centering
        \includegraphics[width=\linewidth]{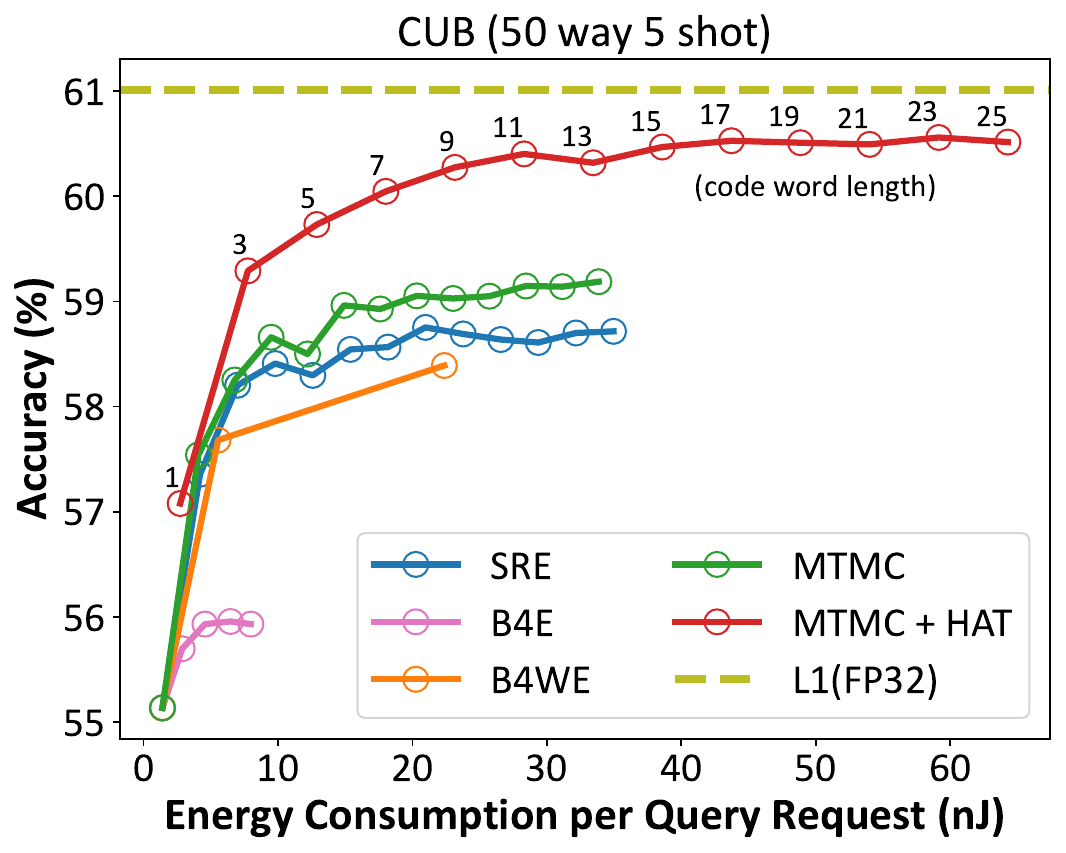}
        \vspace{-2em}
        \caption{}
    \end{subfigure}
    \vspace{-1em}
    \caption{Pareto fronts of energy-accuracy trade-off for (a) Omniglot and (b) CUB datasets.}
    \label{fig::pareto_front}
    \Description{}
    \vspace{-1em}
\end{figure}
To demonstrate the effectiveness of the proposed methods, we implemented the simple repetition encoding (SRE) used in \cite{ref::sapiens}, base-4 encoding (B4E) used in \cite{ref::b4_encoding} and base-4 weighted encoding (B4WE) used in \cite{ref::b4_weighted} using NAND-based MCAM. SRE simply duplicates the support vectors to improve the robustness to the non-idealities of the emerging memory devices. B4E focuses on enhancing the precision through bit-slicing. B4WE balances between robustness and precision through non-uniform repetition based on B4E. B4WE encodes the quantized vectors through B4E and duplicates the $i$-th code word $4^{i-1}$ times, making the impact of the higher (more important) bit on the voting result more significant.

\Cref{fig::pareto_front} shows the Pareto fronts of energy-accuracy trade-off for the Omniglot and CUB dataset. For fair comparison, AVSS is adopted for all encoding method and the comparison between SVSS and AVSS will be discussed in \Cref{sec::comparison_svss_avss}. 
Except for MTMC+HAT, which employed the proposed HAT training mechanism, we used the controller trained with a standard two-stage training flow \cite{ref::meta_baseline} for SRE, B4E, B4WE, and MTMC to evaluate MTMC's effectiveness. The data points on the curves of B4WE represents the code word length of 1, 5, and 21 due to the limited granularity. On the other hand, the data points on the curves of B4E represents the code word length from 1 to 9 since the corresponding quantization level of code word length 9 is $4^9$, which is large enough and close to the floating points implementation. For SRE, MTMC and MTMC+HAT, the data points on the same curves represent the different code word lengths ranging from 1 to 32 for Omniglot and from 1 to 25 for CUB. We also implement prototypical network \cite{ref::prototypical} with $L_1$ metric as the software baselines.


 Our observation reveals that merely duplicating the support vectors twice (the second data points of SRE) resulted in a significant accuracy enhancement of 2.22\% to 3.36\% , highlighting reliability as a crucial factor. While B4E only focuses on enhancing precision, its lower reliability resulted in minimal accuracy improvements. Conversely, both SRE and B4WE demonstrated considerable accuracy gains as the code word length increased due to the repetition technique for improving reliability. However, for B4WE, a bottleneck effect negated the accuracy enhancements derived from the increased precision, resulting in the accuracy slightly lower than SRE. Unlike these methods, the cumulative encoding rule of MTMC takes both precision and reliability into considerations, enabling it to outperform SRE and B4WE with accuracy improvements ranging from 0.34\% to 0.80\%. Furthermore, when applying the controller trained with the proposed HAT method (MTMC+HAT), the pareto front can be pushed further to a better trade-off with $1.25\%$ to $1.8\%$ accuracy improvement with the same energy consumption compared to MTMC.



\subsection{Comparison between SVSS and AVSS}
\label{sec::comparison_svss_avss}
\begin{table}
  \caption{Accuracy and throughput comparison between SVSS and AVSS with HAT}
  \vspace{-1em}
  \label{tab::svss_vs_avss_with_hat}
  \begin{tabular}{|c|cc|cc|}
    \hline
    Dataset & \multicolumn{2}{c|}{\textbf{Omniglot}}   & \multicolumn{2}{c|}{\textbf{CUB}} \\
    \hline
    &\multicolumn{1}{c|}{\textbf{SVSS}}&\textbf{AVSS}&\multicolumn{1}{c|}{\textbf{SVSS}}&\textbf{AVSS} \\
    \hline
    \textbf{Accuracy} (\%) & \multicolumn{1}{c|}{92.27} & 91.31 (-0.96) & \multicolumn{1}{c|}{60.95} & 60.30 (-0.65)\\
    \hline
    \begin{tabular}[c]{@{}c@{}}\textbf{Throughput}\\ ($search/s$)\end{tabular} & \multicolumn{1}{c|}{312.5} & 10000 (32$\times$) & \multicolumn{1}{c|}{40} &  1000 (25$\times$)\\
    \hline
  \end{tabular}
  \vspace{-2em}
\end{table}

To provide a clear understanding of the trade-off between SVSS and AVSS, we conducted a detailed analysis of accuracy and throughput in both scenarios across two datasets. HAT is also incorporated in this experiment. In the implementation of HAT, SVSS employs the standard quantization method, whereas AVSS utilizes an asymmetric quantization scheme as detailed in Section \ref{sec::AVSS}. As demonstrated in \Cref{tab::svss_vs_avss_with_hat}, AVSS significantly enhances the efficiency of the VSS process, shortening the searching process of VSS by $25\times$ to $32\times$, while incurring only minimal accuracy decrements of $0.65\%$ to $0.96\%$. This minimal loss in accuracy is attributed to the effective implementation of HAT, which optimizes the quantization process to align closely with the hardware capabilities, thereby preserving high accuracy levels while boosting throughput.




\section{Conclusion}
In this paper, we introduce an efficient and reliable in-memory search mechanism using NAND-based MCAM for many-class FSL. We propose three methods, including MTMC, AVSS, and HAT, to leverage the NAND-based MCAM's advantages and overcome its constraints effectively. Apart from the prior works, MTMC leverages the large capacity of NAND-based MCAM to increase the precision of the vectors in VSS while mitigating the bottleneck effect of NAND-based MCAM. Moreover, searching iterations are significantly reduced by AVSS, while the accuracy drop caused by non-ideal effects of MCAM are mitigated by HAT. Experimental results show that our methods improve the overall accuracy by $1.58\%$ to $6.94\%$ compared to the encoding method used in prior works. In addition, up to $32\times$ throughput can be achieve through AVSS with only less than $1\%$ accuracy drop compared to SVSS.


\end{document}